# Mining Twitter to Assess the Determinants of Health Behavior towards Human Papillomavirus Vaccination in the United States

Hansi Zhang[1]*, Christopher Wheldon[2]*, Adam G. Dunn[3], Cui Tao[4], Jinhai Huo[5], Rui Zhang[6], Mattia Prosperi[7], Yi Guo[1], Jiang Bian[1]¶

[1]Health Outcomes & Biomedical Informatics, College of Medicine, University of Florida, Gainesville City, FL, United States

[2]National Cancer Institute, Rockville, MD

[3]Centre for Health Informatics, Australian Institute of Health Innovation, Macquarie University, Sydney, NSW 2109, Australia

[4]School of Biomedical Informatics, The University of Texas Health Science Center at Houston, Houston, TX, United States

[5]Department of Health Services Research, Management and Policy, University of Florida, Gainesville City, FL, United States

[6]Department of Pharmaceutical Care & Health Systems, University of Minnesota, Minneapolis, Minnesota, United States

[7]Department of Epidemiology, University of Florida, Gainesville City, FL, United States

* These authors contributed equally to this work

¶ **Corresponding author:** Jiang Bian

2197 Mowry Road, 122, Gainesville, FL 32610; Phone: 352-273-8878; E-mail: bianjiang@ufl.edu

**Keywords:** Twitter, Social media, Human papillomavirus vaccine, Topic modeling, Integrated Behavior Model

# ABSTRACT

**Objectives**

To test the feasibility of using Twitter data to assess determinants of consumers' health behavior towards Human papillomavirus (HPV) vaccination informed by the Integrated Behavior Model (IBM).

**Methods**

We used three Twitter datasets spanning from 2014 to 2018. We preprocessed and geocoded the tweets, and then built a rule-based model that classified each tweet into either promotional information or consumers' discussions. We applied topic modeling to discover major themes, and subsequently explored the associations between the topics learned from consumers' discussions and the responses of HPV-related questions in the Health Information National Trends Survey (HINTS).

**Results**

We collected 2,846,495 tweets and analyzed 335,681 geocoded tweets. Through topic modeling, we identified 122 high-quality topics. The most discussed consumer topic is "*cervical cancer screening*"; while in promotional tweets, the most popular topic is to increase awareness of "*HPV causes cancer*". 87 out of the 122 topics are correlated between promotional information and consumers' discussions. Guided by IBM, we examined the alignment between our Twitter findings and the results obtained from HINTS. 35 topics can be mapped to HINTS questions by keywords, 112 topics can be mapped to IBM constructs, and 45 topics have statistically significant correlations with HINTS responses in terms of geographic distributions.

**Conclusion**

Not only mining Twitter to assess consumers' health behaviors can obtain results comparable to surveys but can yield additional insights via a theory-driven approach. Limitations exist; nevertheless, these encouraging results impel us to develop innovative ways of leveraging social media in the changing health communication landscape.

# BACKGROUND and SIGNIFICANCE

Human papillomavirus (HPV) is the most common sexually transmitted disease (STD) in the United States (US) [1]. Although HPV infections are transient, persistent infection can lead to cancer. An estimated 33,700 new patients are diagnosed with HPV-associated cancers (e.g., anal, penile, cervical, and oral cancers) each year [2] in US. HPV vaccine is effective in preventing most of these HPV-related cancers for individuals in early age [3]. Nevertheless, in 2017, only 48.6% of US adolescents received recommended HPV vaccination series, and 65.5% received ≥1 dose of the series [4]. HPV vaccination coverage also varies greatly by state. Only three states (i.e., District of Columbia: 91.9%, Rhode Island: 88.6%, and Massachusetts: 81.9%) have more than 80% coverage for the first dose, while the bottom three states (i.e., Kentucky: 49.6%, and Mississippi: 49.6%, Wyoming: 46.9%) have coverage rates less than 50% [4]. There is a huge public health needs to increase the awareness of HPV-related issues to promote HPV vaccination.

To increase HPV vaccination initiation and coverage, we first need to understand factors that affect people's health behavior towards vaccination uptake. Recognized by the Integrated

Behavior Model (IBM), a general theory of behavioral prediction, individuals' intention is the most important determinant of their health behaviors (i.e., HPV vaccination uptake in our case), while behavior intention is subsequently determined by attitude (e.g., feelings about the behavior), perceived norms (e.g., the social pressure one feels to perform the behavior), and personal agency (e.g., self-efficacy) [5]. Other factors such as knowledge (i.e., skills to perform the behavior), environmental constraints (e.g., access to care), habits, and salience of the behavior can also directly affect individuals' health behaviors.

Interviews, focus groups, and questionnaires are traditional approaches for understanding these factors that affect individuals' behavior decision-making processes. A few studies used these traditional approaches to examine the determinants of HPV vaccination uptake [6–8]. The rapid growth of social media has transformed the communication landscape not only for people's daily interactions but also for health communication. People want their voices to be heard and voluntarily share massive information about their health history and status, perceived value and experience of care, and many other user-generated health data on social media. A few studies also used social media data to understand individuals' HPV vaccination behavior. Du *et al* leveraged a machine learning-based approach to inspect individuals' attitudes (i.e., positive, neutral, and negative sentiments) about different aspects of HPV vaccination (e.g., safety and costs) using Twitter data [9]. Kein-malpass *et al* mined Twitter data to understand public perception of HPV vaccine through a manual content analysis [10]. Dunn *et al* explored consumer's information exposure related to HPV vaccine on Twitter and found that populations disproportionately exposed to negative topics had lower coverage rates [11]. However, very few studies were guided by any well-established health behavior theories.

Shapiro *et al* used Health Belief Model to code the types of individuals' concerns such as unnecessary (e.g., HPV vaccine is not beneficial), perceived barriers (e.g., perceived harms), cues to action (e.g., influential organizations guiding against HPV vaccine) among many other concerns (e.g., mistrust, undermining religious principles, and undermining civil liberties) [12]. However, they did not compare their social media findings with those obtained from traditional methods (e.g., surveys). The validity of using social media data for understanding behavioral determinants warrants further investigation.

Further, most of these HPV-related social media studies did not consider the different types of users who posted about HPV: 1) those who are involved in health promotion (e.g., government agencies, health organizations, and professionals), and 2) individual consumers discussing policies and their own vaccination experiences. While all forms of HPV information may contribute to the factors that shape vaccination behaviors, distinguishing between promotional information and consumer's discussions may help understand the impact of health promotion on individuals' behaviors.

In this study, we aim to mine Twitter to examine population-level associations between consumers' HPV-related discussions and their vaccination behaviors in the US guided by IBM. We fill three important gaps in prior social media HPV-related studies: 1) we classified HPV-related tweets into promotional information vs. consumers' discussions; 2) we mapped the topics learned from consumers' Twitter discussions to IBM constructs; and 3) we assessed the associations between the learned Twitter topics and responses to HPV-related questions (**Table 4**) in the Health Information National Trends Survey (HINTS) [13] to determine the feasibility of using social media derived measures to match and/or complement survey-based measures of vaccination behaviors. Our study addresses the following research questions (RQs):

1. RQ1: What are the topics discussed in HPV-related tweets?
2. RQ2: Are there any correlations between promotional HPV-related information and consumers' discussions on Twitter in terms of topic distributions?
3. RQ3: Can consumer discussion topics in Twitter be mapped to IBM constructs; and are the geographic distributions of these topics comparable to the determinants measured from HINTS survey?

## METHODS

**Data Sources**

We used three Twitter datasets collected independently using Twitter application programming interface with HPV-related keywords. The three datasets covered overlapping date ranges, spanning from January 2014 to April 2018 (**Table 1**). From a total of 2,846,495 tweets, we removed 248,462 duplicates and retained 2,598,033 tweets.

**Table 1**. The three HPV-related Twitter datasets, their date ranges, keywords used for data collection, and total number of tweets.

| Data Source | Date Range | Keywords[a] | Total Number of Tweets (%)[b] N = 2,598,033 |
|---|---|---|---|
| Collected for | 2016/01 – | cervarix, gardasil, hpv, human | 2,238,433 (86.16%) |

| this project | 2018/04 | papillomavirus | |
|---|---|---|---|
| Dunn *et al*. [11] | 2014/01 – 2016/12 | gardasil, cervarix, hpv + vaccin∗, cervical + vaccin∗ | 423,594 (16.30%) |
| Du *et at*. [14] | 2015/11 – 2016/03 | cervarix, gardasil, hpv, human papillomavirus | 184,468 (7.10%) |

[a] "hpv + vaccin∗" means a tweet has to contain both the word "hpv" and a word starts with "vaccin".

[b] Note that there are overlaps across the three datasets. The percentage indicates the amount of tweets of each dataset over the total number of unique tweets combined.

Further, we obtained survey data from HINTS-4-Cycle-4 (i.e., covering August 2014 to Novemeber 2014) and HINTS-5-Cycle-1 (i.e., January 2017 to May 2017). HINTS is a nationally representative survey on public's use of cancer and health-related information. We extracted responses from 6,962 respondents who answered 8 HPV-related questions from the two datasets. We obtained state-level geographic information and full-sample weight (i.e., to calculate population estimates) of each respondent.

**Data Analysis**

Our data analysis consists of four steps (**Figure 1)** and detailed below.

*Step 1: Data preprocessing*

We first removed non-English tweets using a two-step process. The '*lang*' attribute specifies the langue of the tweet, identified by Twitter's internal language detection algorithm [15]. If the '*lang*' attribute was not available, we used Google's language detection algorithm [16] to identify the language. We also made a few other data cleaning efforts. We removed 1) hashtag symbols ("#"), 2) Uniform Resource Locators (URLs), and 3) user mentions (e.g. "@*username*").

We geocoded each tweet to a US state using a tool we developed previously [17]. Twitter users have three options to attach geographic information to their tweets or profiles: (1) a tweet includes a geocode (Global Positioning System [GPS] latitude and longitude) or a geographic '*place*', if it is posted with a GPS-enabled mobile device or the user chose to tag it with a '*place*'; (2) the associated user profile can be geocoded (either to a GPS location or a '*place'*); and (3) the user can fill the '*location*' attribute with free-text [18]. If geocodes were available, we attempted to resolve the locations through reverse geocoding using GeoNames [19], a public geographical database. Very few (i.e., 0.85% of all tweets) tweets have geocodes [20]. For most tweets, we matched the free-text '*location*' with lexical patterns indicating the location of the user such as a state name (e.g., "*Florida*") or a city name in various possible combinations and formats (e.g., "——, fl" or "——, florida").

*Step 2: Rule-based categorization of the tweets*

Previously, we built classifiers to filter out irrelevant tweets [21, 22]. Nevertheless, in some cases [23], the keywords used for data collection were specific enough; thus, very few tweets were irrelevant. We randomly annotated 100 tweets and found that only 2 tweets were irrelevant to HPV (i.e., 98% were relevant). We thus considered all our tweets as HPV-related without needing a complex classifier.

We then categorized these tweets into either: (1) promotional information, or (2) consumers' discussions. Consistent with our previous findings [22], tweets that contain URLs are more likely to be promotional information, where the URLs are links to HPV-related news, research findings, and health promotion activities. We randomly annotated 100 tweets with URLs and found 95% are promotioinal information. Further, users can "*quote*" another tweet or online resources (e.g., a web page) but with additional comments expressing their own opinion, and the original quoted tweets (or web pages) are converted into URLs. Twitter users can also "*retweet*" another tweet (i.e., starts with "*rt*"); nevertheless, the original tweet is not converted into a URL (but URLs in the original tweet were preserved). Based on these observations, we devised a set of simple, yet effective rules as shown in **Figure 2**. Note that these rules were applied on the original tweets before removing URLs.

*Step 3: Topic modeling*

Topic modeling, a statistical natural language processing (NLP) approach, is wildly used for finding abstract underlying topics in a collection of documents. We used the latent Dirichlet allocation (LDA) model [24] to extract topics from our HPV-related tweets. In LDA, each document (i.e., a tweet) is modeled as a mixture of topics, and each topic is a probability distribution over words. The LDA algorithm exploits documental-level word co-occurrence patterns to discover underlying topics. Based on a prior study, we first removed stop words (e.g., "*the*", "*a*") and words that occurred ≤3 times in our corpus [25].

Even though LDA is an unsupervised approach, the number of topics needs to be set *a priori*. We tested three statistical methods to find the best number of topics: (1) Arun2010 [26], (2)

Cao2009 [27], and (3) Deveaud2014 [28]. However, these methods did not converge on our Twitter corpus. One possible reason is that Twitter messages are short but the number of tweets is huge; thus, we may need a large number of topics to obtain a reasonable model [29]. Thus, we choose a relatively large number (i.e., 150 topics) based on parameters used in similar Twitter LDA studies [29, 30]. We also visualized each topic using the top 10 words as a word-cloud, where the size of each word is proportional to its probability in that topic.

The nature of LDA allows all topics (derived from the entire collection of tweets) to occur in the same tweet with different probabilities, while topics with low probabilities might not actually exist. Thus, we needed to determine a cutoff probability value to select the most representative and adequate topics. We tested a range of cutoff values and manually evaluated a random sample of tweets (i.e., 100) for each tested cutoff value to determine whether the topics (whose probabilities were larger than the cutoff) assigned to each tweet were correct. We selected the lowest cutoff where more than 80% of topic assignments were adequate. After assigning topics for each tweet, we manually evaluated each topic's word-cloud and a sample of associated tweets to determine the topic's: 1) theme, and 2) quality (i.e., a topic was of low quality if more than half of the sample tweets were not relevant to the assigned topic or the word-cloud words do not have a consistent theme).

*Step 4: Research questions*

For RQ1, to identify popular topics, we calculated the percentage of each topic's tweet volume for both promotional information and consumers' discussions w.r.t. the total number of tweets within each category, based on which we ranked the topics.

For RQ2, to assess correlations between promotional information and consumers' discussions, we calculated the Pearson correlation coefficient between the two in terms of monthly tweet volumes for each topic.

For RQ3, we first mapped high-quality topics (**Step 3**) directly to IBM constructs through manually examining each topic's word-cloud and a sample of 10 associated consumer tweets (promotional information does not reflect thoughts from lay consumers, thus not considered) by two annotators (HZ and JB). For example, a twitter topic "*HPV related cancers*" with a sample consumer tweet—"*HPV is a contributor to the rise in mouth cancer…*"—can be mapped to the "*knowledge" construct* in IBM. A topic is excluded if it does not represent consumers' discussions (i.e., more than 5 consumer tweets are irrelevant to the topic theme). Conflicts between the two annotators are resolved through discussions with a third reviewer (YG).

We also mapped the high-quality topics to HPV-related HINTS questions. To do so, we first grouped similar HINTS questions into question groups (QGs) and mapped the QGs to IBM constructs. For example, questions "*…HPV can cause anal cancer?*" and "*…HPV can cause oral cancer?*" can be grouped into a QG *"Knowledge on HPV-cancer relationships"* to the "*knowledge*" construct. We then manually extracted key terms from each survey question and mapped topics to the question based on matching these keywords with the top 20 words in each Twitter topic. For example, "*HPV*", "*oral*" and "*cancer*" were extracted from "*…HPV can cause oral cancer?*" and can be mapped to topic-81 "*HPV and oral cancer*", where the top 5

keywords are "*oral*", "*cancer*", "*hpv*", "*sex*", and "*dentist*". A topic is mapped to a QG if it is mapped to any questions in the group.

To assess whether Twitter data are comparable to survey in measuring the determinants of vaccination behavior guided by IBM, our first step is to establish the correlations between consumer-related HPV topics and population estimates derived from HPV-related HINTS questions at state level. To do so, we first aggregated geocoded tweets of the same state and derived the normalized geographic distribution of each topic at the state level (i.e., divided the number of tweets for each topic by the total number of consumer tweets in a state). From survey data, to obtain the normalized geographic distribution of HINTS responses, we divided the number of respondents with the answers of interest (e.g., responded "*Yes*" to "*…HPV can cause anal cancer?*" indicating the respondent has the "knowledge") by the total number of respondents for each state considering each respondent's full-sample weight [31] in HINTS.

After normalized both Twitter and survey data, we calculated the Spearman's rank correlations between Twitter topics and the population estimates (derived from HINTS survey responses to each QG) in terms of their geographic distributions. Note that, considering that we grouped survey questions into QGs, we also combined answers for all questions in that QG (i.e., if the respondent responds with the interested answer for any question in that QG).

## RESULT

**Step 1: Preprocessing**

We removed 958,483 non-English tweets and retained 2,598,033, out of which 335,681 (12.92%) tweets could be geocoded to a US state for further analysis.

**Step 2: Rule-based categorization of the tweets**

We annotated 100 random tweets and assessed the performance of our rules, which achieved a precision of 84.21%, a recall of 86.00%, and a F-measure of 85.10%. We applied these rules on all the geocoded tweets. Out of the 335,681 geocoded tweets, 93,693 (27.91%) tweets were classified as consumers' discussions and 241,988 (72.09%) tweets were promotional information.

**Step 3: Topic modeling**

We determined the cutoff probability for topic assignment is 0.15, where 84% of 100 randomly selected tweets' topic assignments were adequate. We were able to assign topics to 86.85% (i.e., 291,551) of the geocoded tweets. We manually evaluated each topic's word-cloud and 10 random associated tweets to determine its quality, eliminated 28 low-quality topics (out of 150), and considered the remaining 122 topics (i.e., associated with 281,712 tweets) in further analyses. **Table 2** shows example tweets and the topics associated with each tweet.

**Table 2.** Example tweets and associated topics.

| Tweet[a] | Top 3 Topics (%)[b] |
|---|---|
| "RT @user1: @ user2 they have had a rise in Anal Cancers due to HPV virus and the fact they think anal sex | Topic-11 "*pap smear test*" (19%) <br><br> Topic-14 "*HPV related cancers*" (13%) <br><br> Topic-106 "*cervical cancer and death*" (11%) |
| *maintains virginity"* | |
| *"The startling rise in oral cancer in men - another good reason to vaccinate males against HPV https://t.co/xxx"* | Topic-81 *"HPV and oral cancer"* (20%) <br> Topic-14 *"HPV related cancers"* (14%) <br> Topic-147 *"doctors' discussions of vaccine"* (14%) |
| *"I'm making health calls: HPV infection can cause penile cancer in men; and anal cancer, cancer of the back of the throat."* | Topic-14 *"HPV related cancers"* (29%) <br> Topic-59 *"HPV causes cancer"* (11%) <br> Topic-18 *"HPV, HPV vaccination and HPV related cancer for man"* (8%) |
| *"RT @user1: You don't have to have sex to get an STD. Skin-to-skin contact is enough to spread HPV. https://t.co/xxx"* | Topic-24 *"STD and cervical cancer cure"* (39%) <br> Topic-56 *"questions about HPV and vaccine"* (12%) <br> Topic-64 *"sex and HPV vaccine"* (11%) |
| *"Please join us for a Facebook event about cervical cancer treatments on 1/26 at 2:00 pm ET https://t.co/xxx htttps://t.co/xxx"* | Topic-114 *"cervical cancer treatment"* (20%) <br> Topic-67 *"cervical cancer diagnosis and signs"* (11%) <br> Topic-95 *"treatment"* (9%) |
| *"HPV Vaccine That Helps Prevent Cervical Cancer in Women May Cut Oral Cancer https://t.co/xxx"* | Topic-106 *"cervical cancer and death"* (29%) <br> Topic-0 *"blogs about HPV and HPV commercials"* (14%) <br> Topic-27 *"HPV infection"* (10%) |
| *"my doctor accidentally gave me a fourth dose of gardasil so thats where i'm at today"* | Topic-147 *"doctors' discussions of vaccine"* (18%) <br> Topic-138 *"discussions about HPV and vaccine"* (15%) |

|  | Topic-37 "*cervical cancer risk*" (11%) |
| --- | --- |
| "*New CDC Recommendations for HPV Vaccines https://t.co/xxx*" | Topic-123 "*vaccine recommendation*" (25%) |
|  | Topic-85 "*HPV vaccine related needs*" (14%) |
|  | Topic-2 "*HPV shot / Gardasil shot*" (11%) |

[a] These tweets are slightly altered to preserver the privacy of the Twitter users without changing the meaning of the original tweets.

[b] Topics and associated probability. Note that the cutoff probability is set to 0.15; thus, topics with <0.15 probabilities were eliminated for each tweet.

## Step 4: Research questions

*RQ1: What are the topics discussed in HPV-related tweets?*

The top 3 topics are visualized as word-clouds in **Figure 3**.

*RQ2: Are there any correlations between promotional HPV-related information and consumers' discussions on Twitter in terms of topic distributions?*

We plotted the monthly tweet volumes of both promotional information and consumers' discussions in **Figure 4**. 87 out of the 122 high-quality topics are correlated between promotional information and consumers' discussions ($p < 0.05$) in terms of their monthly volumes. The top 10 correlated topics are presented in **Table 3**.

**Table 3.** Pearson correlation coefficients between promotional information and consumers' discussions in terms of each topic's monthly tweet volumes.

| Topic | Correlation | P-value | Tweet Volumes[a] |
| --- | --- | --- | --- |

|  | Coefficient |  | # of Promotional Information Tweets (%) | # of Consumers' Discussions Tweets (%) |
|---|---|---|---|---|
|  |  |  | N = 241,988 | N = 93,693 |
| Topic-5 "*pap smear test*" | 0.9517 | < 0.01 | 5,598 (2.31%) | 2,331 (2.49%) |
| Topic-89 "*cervical cancer awareness month*" | 0.9252 | < 0.01 | 4,614 (1.98%) | 1,854 (1.91%) |
| Topic-103 "*knowledge of HPV and cervical cancer facts*" | 0.8758 | < 0.01 | 5,622 (2.32%) | 1,678 (1.79%) |
| Topic-65 "*cervical cancer in black women*" | 0.8096 | < 0.01 | 5,535 (2.29%) | 2,975 (3.18%) |
| Topic-75 "*cervical cancer screening*" | 0.7625 | < 0.01 | 8,628 (3.57%) | 12,500 (13.34%) |
| Topic-117 "*cervical and breast cancer*" | 0.7608 | < 0.01 | 2,487 (1.03%) | 1,498 (1.60%) |
| Topic-106 "*cervical cancer and death*" | 0.7500 | < 0.01 | 6,896 (2.85%) | 2,772 (2.96%) |
| Topic-14 "*HPV-related cancers*" | 0.7070 | < 0.01 | 4,853 (2.01%) | 1,615 (1.72%) |
| Topic-59 "*HPV causes cancer*" | 0.5247 | < 0.01 | 10,649 (4.40%) | 3,961 (4.23%) |
| Topic-45 "*HPV vaccine in* | 0.4506 | < 0.01 | 8,320 (3.44%) | 1,915 (2.04%) |

| boys and girls" | | | | |

<sup>a</sup> For clarity, we only presented top 10 correlated topics with tweet volumes greater than 1,000. For volume less than 1,000, the sample size might be too small to justify the *correlation* even with p < 0.05.

***RQ3: Can consumer discussion topics in Twitter be mapped to IBM constructs; and are the geographic distributions of these topics comparable to the determinants measured from HINTS survey?***

We found 112 out the 122 high-quality topics are relevant to consumers' discussions and can be mapped to 6 different IBM constructs (**Figure 5**): 1) "*feelings about behavior*" (97 topics), 2) "*behavioral beliefs*" (92 topics), 3) "*normative beliefs - other's behavior*" (36 topics), 4) "*knowledge*" (23 topics), 5) "*normative beliefs - other's expectation*" (7 topics), and 6) "*environmental constrains*" (2 topics). Note that a topic can be mapped to multiple IBM constructs. The inter-rater reliability between the two annotators is 0.78.

We grouped 8 HPV-related HINTS questions into 5 QGs and mapped the 5 QGs to 3 types of IBM constructs as shown in **Table 4**. Out of the 122 topics, 35 topics were mapped to HINTS questions based on keyword-matching through manual review (i.e., kappa = 0.72).

We then explored two sets of Spearman's rank correlations between the geographic distributions of 1) the 35 Twitter topics mapped to HINTS QGs based on keyword matching; and 2) the 112 topics mapped directly to IBM constructs, and the population estimates derived from HINTS data as shown in **Table 4**. **Figure 6** shows an example of three geographic heatmaps for: 1) HINTS QG2, 2) topic-17 that was mapped to QG2 by keywords with a low correlation ($\rho$: 0.35;

p < 0.05), and 3) topic-127 that was not mapped to QG2 by keywords but had the strongest correlation with QG2 ($\rho$: 0.55; p < 0.01).

**Table 4.** Mapping topics in consumers' discussions to the HPV-related survey questions in HINTS and corresponding constructs in the Integrated Behavior Model.

| HPV-related survey questions in HINTS | # of mapped topics | Integrated Behavior Model Construct | Correlation for the 35 topics mapped to HINTS questions by keywords[a] | Correlation for the 112 topics mapped to IBM constructs through manual review (Top 3)[b] |
|---|---|---|---|---|
| QG1. Knowledge on HPV-cancer relationships:<br>   a. Do you think HPV can cause anal cancer?<br>   b. Do you think HPV can cause oral cancer?<br>   c. Do you think HPV can cause penile cancer? | 3 | Knowledge | Topic-81 "*HPV and oral cancer*" ($\rho$: 0.29; p < 0.05) | Topic-111 "*HPV vaccine and vaccine mandate*" ($\rho$: 0.62; p: < 0.01)<br><br>Topic-127 "*HPV symptom and vaccine*" ($\rho$: 0.58; p < 0.01)<br><br>Topic-124 "*vaccine saves lives*" ($\rho$: 0.50; p < 0.01) |
| QG2. Do you think that HPV is a sexually transmitted disease (STD)? | 6 | Knowledge | Topic-17: "*HPV, service and HPV transmission*" ($\rho$: 0.35; p < 0.05) | Topic-127 "*HPV symptom and vaccine*" ($\rho$: 0.55; p < 0.01)<br><br>Topic-23 "*HPV epidemics*" ($\rho$: 0.40; p < 0.01)<br><br>Topic-74 "*HPV, vaccine cost and impact*" ($\rho$: 0.35; p <0.01) |
| QG3. Do you think HPV requires medical treatment or will it usually go away on its own without treatment? | 2 | Knowledge | No statistically significant topics | Topic-149 "*vaccine victims*" ($\rho$: 0.44; p: < 0.01)<br><br>Topic-9 "*fight cervical cancer and hpvvax*" ($\rho$: 0.35; p < 0.01)<br><br>Topic-43 "*early detection of cervical* |

| | | | | *cancer*" (correlation: 0.27; p < 0.01) |
|---|---|---|---|---|
| QG4. In your opinion, how successful is the HPV vaccine at preventing cervical cancer? | 26 | Attitude | Topic-55 "*HPV vaccine prevents cervical cancer*" ($\rho$: 0.35, p < 0.01)<br><br>Topic-5 "*Pap smear test*" ($\rho$: 0.32; p < 0.05)<br><br>Topic-112 "*HPV vaccine protects against cancer*" ($\rho$: 0.28; p < 0.05) | Topic-27 "*HPV infection*" ($\rho$: 0.42; p <0.01)<br><br>Topic-4 "*cervical cancer and Andrew's story*" ($\rho$: 0.41; p <0.01)<br><br>Topic-146 "HPV, vaccine and *sexual behavior*" ($\rho$: 0.37; p: < 0.01) |
| QG5. Physician recommendation of HPV vaccination:<br>   a. In the last 12 months, has a doctor or health care professional ever talked with you or an immediate family member about the HPV shot or vaccine?<br>   b. In the last 12 months, has a doctor or health care professional recommended that you or someone in your immediate family get an HPV shot or vaccine? | 2 | Perceived Norm | No statistically significant topics | Topic-91 "*cervical cancer rate and vaccination rate*" ($\rho$: 0.44; p: < 0.01)<br><br>Topic-88 "*HPV vaccine and public health*" ($\rho$: 0.36; p < 0.01)<br><br>Topic-27 "*HPV infection*" ($\rho$: 0.32; p < 0.01) |

[a] Only topics that have significant correlations (p < 0.05) with the survey question groups are listed.

[b] 112 out of the 122 high-quality topics were mapped to IBM constructs regardless of whether the topic can be mapped to the survey question group or not.

## DISCUSSION

In this study, we explored whether user-generated content on Twitter can be used to assess determinants of health behavior, which are traditionally measured through survey questions. We used methods such as topic modeling on HPV-related tweets to answer our three research questions.

For RQ1, we found that the most popular HPV-related topics among consumers on Twitter are "*cervical cancer screening*" and "*defunding of planned parenthood*", which account for 24.92% of all consumers' tweets. The topic "*defunding of planned parenthood*" is also related to "*cervical cancer screening*", as planned parenthood provides 281,063 Papanicolaou tests for cervical cancer screening each year [32]. Further, the popular topics are similar between promotional information and consumer's discussions, where 5 out of the top 10 topics are the same across the two.

For RQ2, we found that 87 consumer topics (out of 122) are correlated with promotional information, suggesting that promotional health information on Twitter certainly has an impact on consumers' discussions, which is consistent with our previous study on Lynch syndrome [22]. These strong correlations might, from another perspective, indicate that coordinated national efforts and promotion strategies on raising public awareness of HPV have been rather successful in recent years.

In RQ3, for the 35 topics mapped to HINTS questions by keywords, most of these topics have a negligible correlation (i.e., <= 0.3) with HINTS data. One of the two highest correlations we found is between topic-55 "*HPV vaccine prevents cervical cancer*" and QG4: "*how successful is HPV vaccine at preventing cervical cancer?*" ($\rho$: 0.35, $p < 0.01$). One potential reason for these low correlations is that the topics learned using LDA can contain multiple themes (e.g., topic-24 "*STD and cervical cancer*" contains two themes: "*STD*" and "*cervical cancer*". On the other hand, each survey item in HINTS only measures a specific theme (e.g., topic-24 was mapped to QG2: "*Do you think that HPV is a sexually transmitted disease (STD)?*"). Thus, the tweets related to the themes that were not captured in the survey question (e.g., "*cervical cancer*" in this case) are "noises" that lead to a biased correlation measure. To assert the "*true*" correlations, a method that can further separate each topic into sub-themes is needed. Further, depending on what the survey question measures, merely counting the number of tweets in the topic may not yield an accurate measure of the correlation. For example, for survey questions that measure attitude, counting only the tweets that express "*attitudes*" using sentiment analysis may yield better results.

Furthermore, topics emerged from tweets may provide more insights towards understanding individuals' attitude and beliefs about HPV vaccination, which are important predictors of their health behavior (i.e., HPV vaccination uptake). In **Figure 5**, topic-14 "*HPV-related cancers*" is mapped to the question in QG1: "*Can HPV cause oral cancer?*", where from its word-cloud, we not only found words related to "*oral cancer*" (e.g., "*throat cancer*"), but also keywords related to other cancers (e.g., "*penile cancer*"). Through examining tweets from that topic, we found examples where users are linking HPV to not only oral cancer but also other types of cancer (e.g., "*I'm making health calls: HPV infection can cause penile cancer in men; and anal cancer.*").

For the 112 topics mapped manually to IBM constructs, 45 topics are correlated ($p < 0.05$) with HINTS responses: 11 with negligible correlations (i.e., $\rho < 0.3$), 30 with low correlations (i.e., $0.3 < \rho < 0.5$), and 4 topics with moderate correlations (i.e., $0.5 < \rho < 0.7$). Most of these topics are related to people's "*attitude*" and "*perceived norm*". However, constructs such as "*personal agency*", "*habit*" and "*salience of the behavior*" are not found in these topics. One possible reason is that compared with "*attitude*" and "*perceived norm*", "*personal agency*" are more difficulty to identify. People may be more willing to talk about their feelings and perceived norms (e.g., other's behavior about HPV vaccination) than their own self-efficacy issues in performing the behavior.

In addition, we found that highly correlated topics and HINTS QGs do not necessarily belong to the same IBM constructs. For example, topics that have high correlations with "*knowledge*" related HINTS questions are all mapped to the construct "*attitude*". These are not necessarily "wrong" results. For example, topic-4 "*cervical cancer and Andrew's story*" (mapped to "normative beliefs") are highly correlated with QG4 (mapped to "*attitude*"). A possible explanation is that the discussion of Andrew's behavior in fighting cervical cancer can be considered as "*normative beliefs – other's behavior*", which will impact people's attitude [33].

## LIMITATIONS

First, social media users are different from the general population. Twitter users are younger than the general population [34]. Thus, the representativeness of social media populations should be carefully considered when interpreting study findings. The presence of bots and fake accounts may also distort the representativeness of our findings. Further, the keywords used across the three datasets are slightly different, which may lead to data selection bias.

Second, sampling units in social media studies (e.g., tweets) are different from traditional survey research (e.g., individuals). In Twitter, a user can have multiple relevant posts, and even multiple accounts. Measures derived from counting tweets might be different from surveys that count individual respondents.

# CONCLUSION

Not only mining Twitter to assess consumers' health behaviors can obtain results comparable to surveys but can yield additional insights via a theory-driven approach. An adequate understanding of the inherent limitations in social media data is always important. Nevertheless, these encouraging results impel us to further develop innovative ways of 1) using social media data (e.g., to understand factors that are precursors to adopting a health behavioral change), and 2) leveraging social media platforms (e.g., to design creative and tailored intervention strategies).

# FUNDING

This work was supported in part by NIH grants UL1TR001427, the OneFlorida Clinical Research Consortium (CDRN-1501-26692) funded by the Patient Centered Outcomes Research Institute (PCORI), and NSF Award #1734134. The content is solely the responsibility of the authors and does not necessarily represent the official views of the NIH, PCORI or NSF.

# COMPETING INTERESTS

The authors have no competing interests to declare.

# CONTRIBUTORS

The work presented here was carried out in collaboration among all authors. JB, and CW designed study. JB, AD, and CT were involved in acquisition of the data. HZ carried out the experiments. JB and HZ wrote the initial draft of the manuscript. AD, CT, RZ, YG, MP and JH edited the manuscript and provided critical feedback.

**Figures and figure legends**

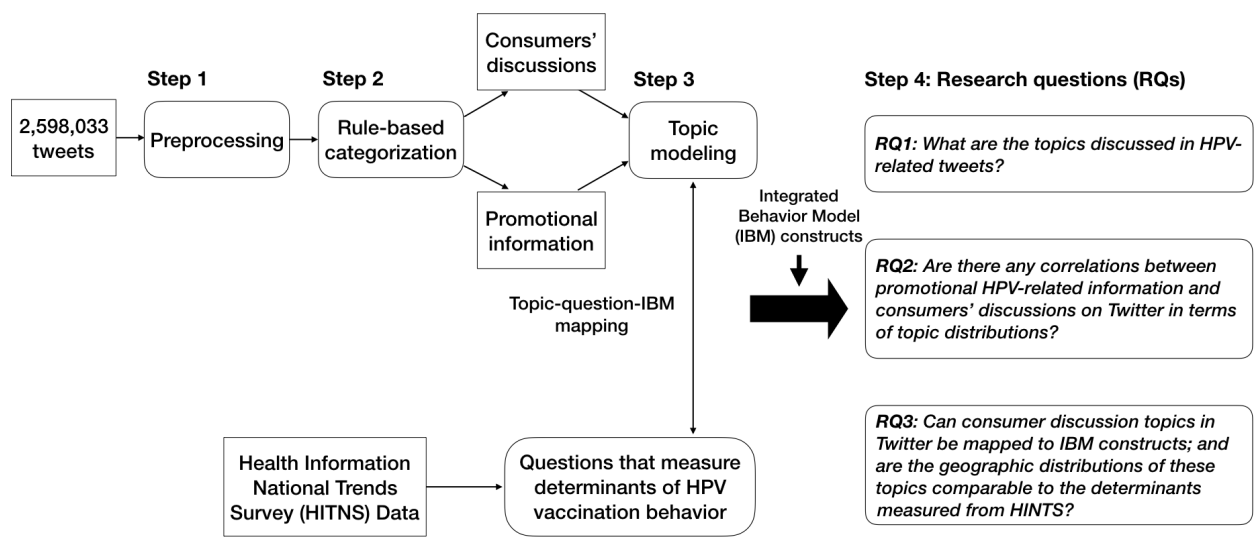

**Figure 1**. The overall data analysis workflow. *Our analysis consists of four steps: (1) data preprocessing; (2) rule-based classification of the tweets into either promotional information or consumers' discussions; (3) applying topic modeling to discover major discussion themes and exploring associations between topics in consumers' Twitter discussions and responses to the 8 HPV-related HINTS questions; and 4) based on these analyses, answering three RQs.

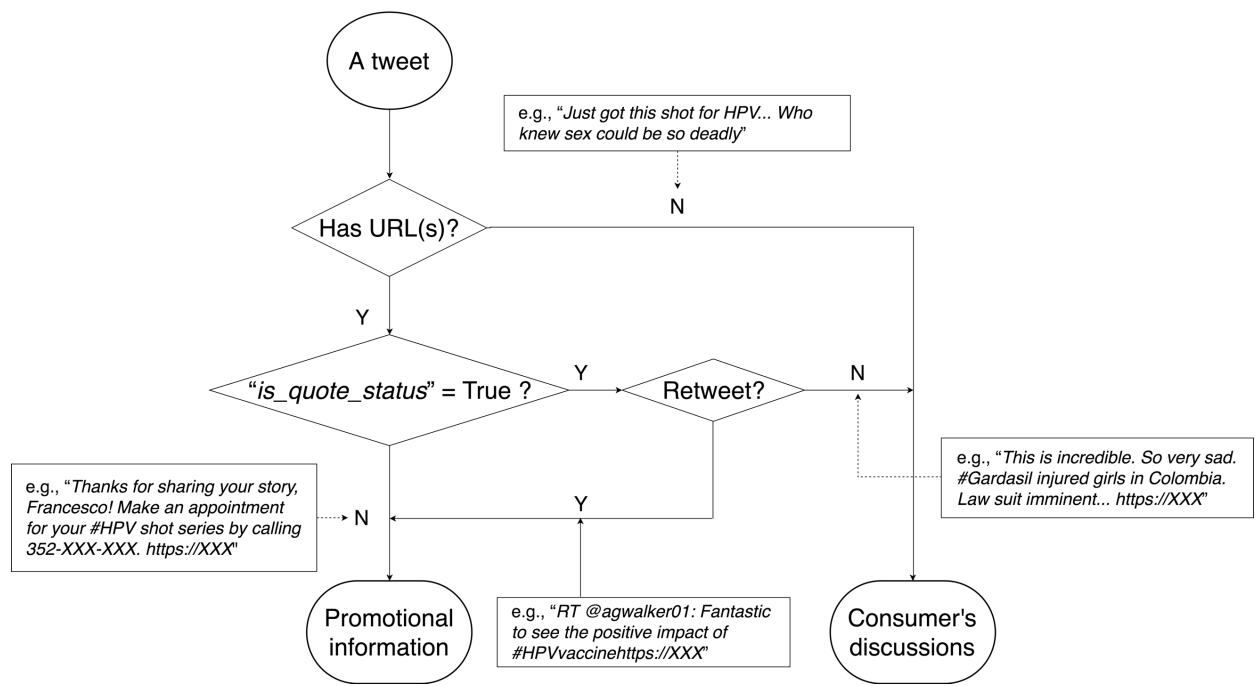

**Figure 2.** A rule-based categorization of the tweets into promotional HPV-related information and consumers' discussions. *If a tweet does not include a URL, it is considered as a consumer discussion. Even if it is a retweet (i.e., starts with 'rt'), the retweet is consumers' discussions as we considered the user who retweeted agrees with the original user discussion and the original tweet is also consumers' discussions (since there is no URLs). When a tweet contains URLs, the rules are more complex: 1) if a tweet is quoting another tweet or web resources (i.e.,

"*is_quote_status*" = True) and not a retweet, it is considered as consumers' discussions. In the special case where the tweet is a retweet of a quoting tweet, we consider this as promotional information because we are unable to determine which of the comments the current user agrees with. In essence, when a tweet is a retweet, we classified the retweet based on the original tweet; and 2) if a tweet is not a quoting tweet, it is considered as promotional information.

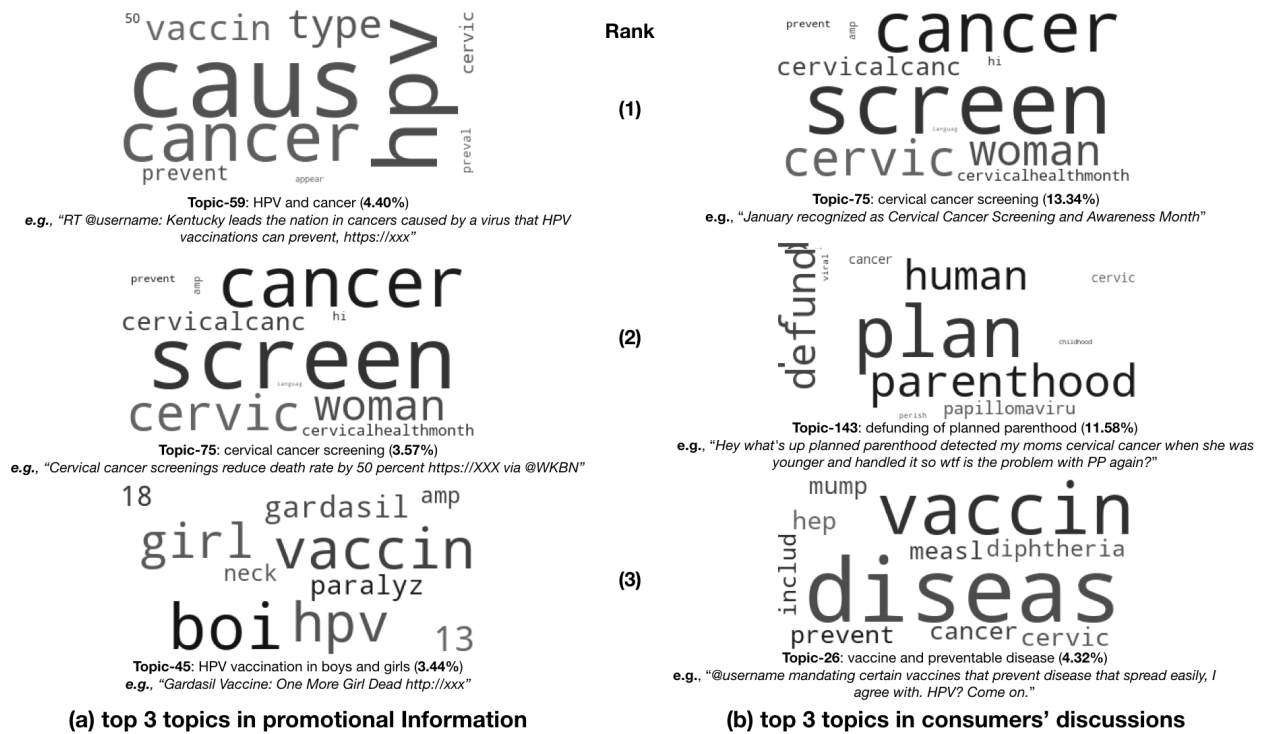

**Figure 3.** The 3 most popular topics in (a) promotional information and (b) consumers' discussions related to HPV and HPV vaccination.

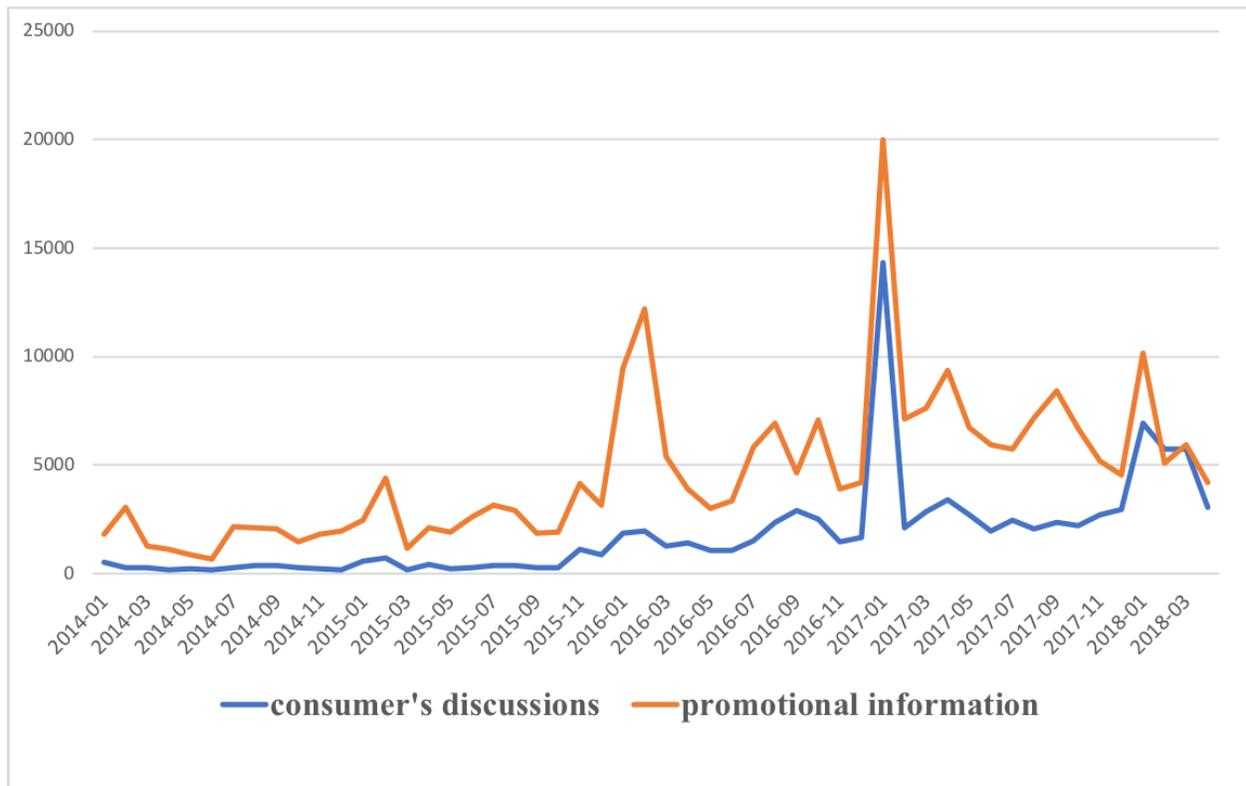

**Figure 4.** The monthly tweet volumes of promotional HPV-related information and consumers' discussion.

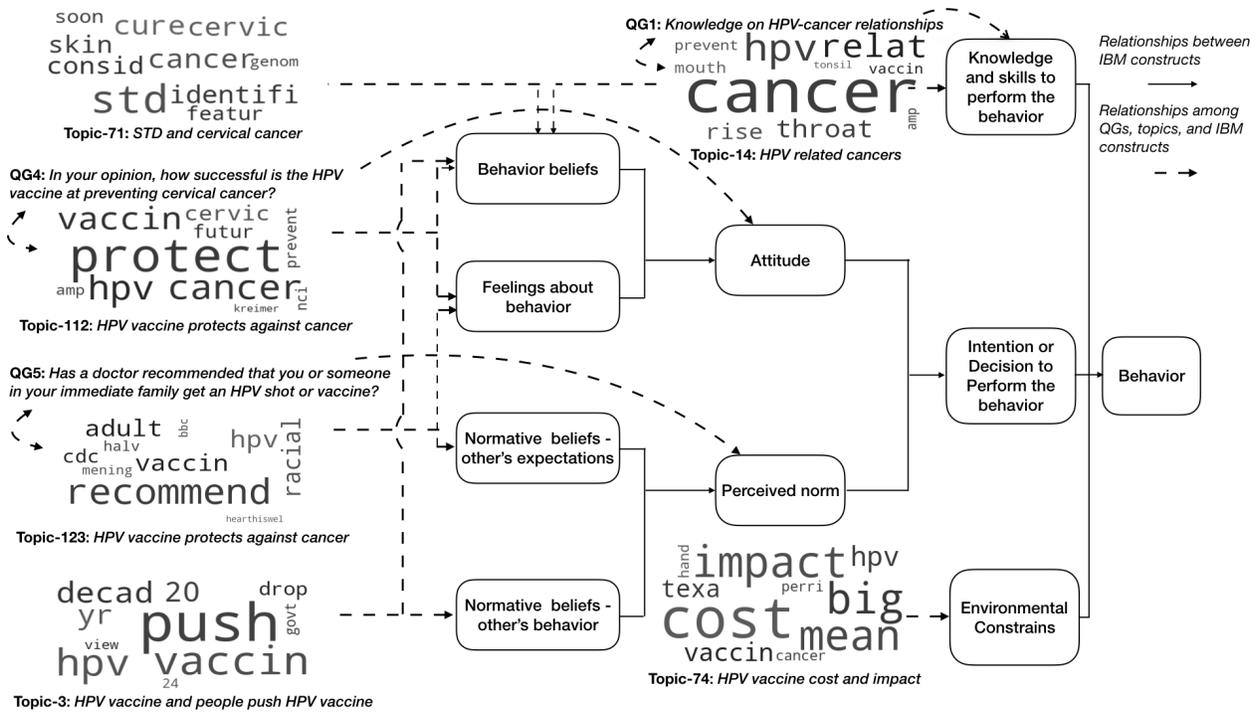

**Figure 5.** Mapping consumer discussion topics to constructs in the Integrated Behavior Model (IBM), including both 1) topics directly mapped to IBM constructs, and 2) topics first mapped to question groups (QGs) and then mapped to IBM constructs (e.g., knowledge – QG1, attitude QG4, perceived norm – QG5).

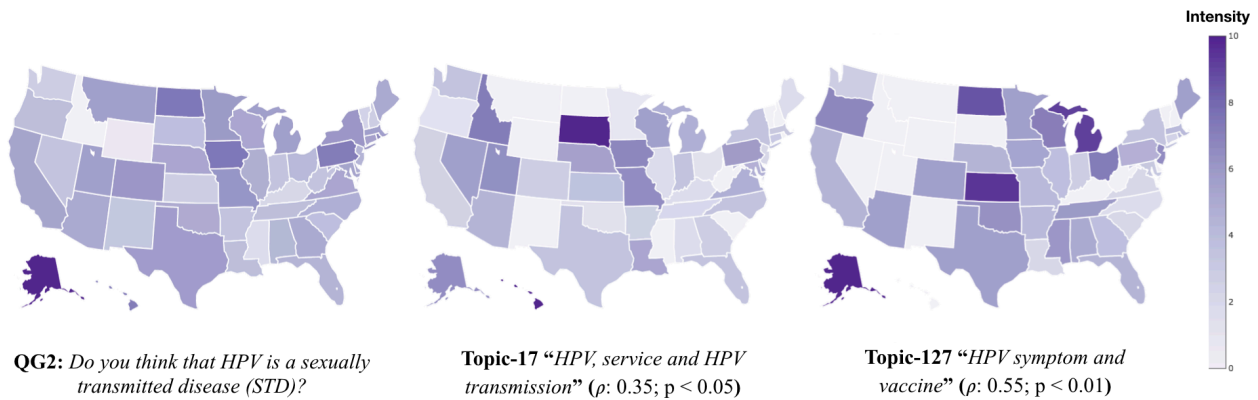

**QG2:** *Do you think that HPV is a sexually transmitted disease (STD)?*

**Topic-17** "*HPV, service and HPV transmission*" ($\rho$: 0.35; p < 0.05)

**Topic-127** "*HPV symptom and vaccine*" ($\rho$: 0.55; p < 0.01)

**Figure 6.** Geographic heatmaps for the state-level distributions of : 1) the responses to HINTS QG2, 2) the number of tweets in topic-17 that was mapped to QG2 by keywords with a correlation $\rho$: 0.35 (p < 0.05), and 3) the number of tweets in topic-127 that was NOT mapped to QG2 by keywords but had the strongest correlation with QG2 ($\rho$: 0.55; p < 0.01). The intensity of the color is proportional to the volumes of tweets assigned to that topic or the number of HINTS responses of interest.